\documentclass[10pt]{article}
\usepackage{amsfonts}
\usepackage{pstricks}
\usepackage{pst-node}
\usepackage{epsfig}
\usepackage{epsfig}
\usepackage[latin5]{inputenc}
\usepackage[T1]{fontenc}
\usepackage{lipsum}
\usepackage{amsmath}
\usepackage{authblk}

\begin{document}
                \def\ba{\begin{eqnarray}}
                \def\ea{\end{eqnarray}}
                \def\w{\wedge}
                \def\d{\mbox{d}}
                \def\D{\mbox{D}}

\begin{titlepage}
\title{Weyl Neutrinos In Plane Symmetric Spacetimes}
\author{Tekin Dereli${}^{1,2,}$\footnote{tdereli@ku.edu.tr, tekindereli@maltepe.edu.tr} , Yorgo \c{S}eniko\u{g}lu${}^{1,}$\footnote{yorgosenikoglu@maltepe.edu.tr}}
\date{%
    ${}^{1}$ \small Department of Basic Sciences, Faculty of Engineering and Natural Sciences, \\Maltepe University, 34857 Maltepe, \.{I}stanbul, Turkey\\%
    ${}^{2}$ \small Department of Physics, Ko\c{c} University, 34450 Sar{\i}yer, Istanbul,Turkey\\[2ex]%
    \today
    }
\maketitle



\begin{abstract}
We investigate complex quaternion-valued exterior differential forms over 4-dimensional Lorentzian spacetimes and explore Weyl spinor fields as minimal left ideals within the complex quaternion algebra. The variational derivation of the coupled Einstein-Weyl equations from an action is presented, and the resulting field equations for both first and second order variations are derived and simplified. Exact plane symmetric solutions of the Einstein-Weyl equations are discussed, and two families of exact solutions describing left-moving and right-moving neutrino plane waves are provided. The study highlights the significance of adjusting a quartic self-coupling of the Weyl spinor in the action to ensure the equivalence of the field equations.
 \end{abstract}

\vskip 1cm


\end{titlepage}

\newpage

\section{Introduction}

Neutrinos are weakly interacting subatomic particles that pervade the universe \cite{close}. There are three families of neutrinos that have long been objects of scientific inquiry due to their elusive nature and potential implications for our understanding of the cosmos. Produced in a range of astrophysical sources, including the sun, supernovae, and active galactic nuclei, neutrinos offer unique insights into the workings of these objects and their evolution in space and time.
Moreover, neutrinos have strong implications for the fundamental laws of physics\cite{senjanovic}. The discovery of neutrino oscillations - the phenomenon where neutrinos transform from one flavor to another as they travel through space - has revealed that neutrinos have mass. This in turn implies modifications of the Standard Model of particle physics, where neutrinos were assumed massless.
The  study of neutrinos is an active area of research.  Physicists employ a variety of experimental techniques; constructing sensitive neutrino detectors  for use at high-energy particle accelerators and as well as for astrophysical observations. Furthermore computer simulation techniques  are also  widely used.
For example, neutrinos play a crucial role in the evolution of astrophysical objects, such as the sun and supernovae. In the sun, neutrinos are produced in the nuclear fusion reactions that power the star. These neutrinos are sensitive to the temperature, density, and composition of the sun's interior and can be used as probes to study its structure and evolution. In supernovae, neutrinos play a vital role in the explosion mechanism and the synthesis of heavy elements. The study of neutrinos in astrophysical sources has become an essential tool for understanding the universe's most energetic and exotic phenomena.
However, studying neutrinos in astrophysical sources is challenging due to their weak interaction with matter. Only a small fraction of the neutrinos produced in astrophysical sources can be detected, and even those that are detected provide limited information about their sources. Furthermore, neutrinos can be affected by other sources of noise, such as cosmic rays, that can make their detection even more challenging.
One area of active research involves the study of neutrinos in curved spacetimes
where deeper insights into the fundamental properties of neutrinos could be gained \cite{brill-wheeler,wainwright,trim-wainwright,griffiths1,madore}.

In particular, in what follows, we study the dynamics of  Weyl neutrinos  in plane symmetric spacetimes \cite{taub,taub2,taub3}.
 The simplicity of such spacetimes due to their assumed isometries allows for a deeper understanding of the physical processes that occur in neutrino interactions with matter and gravity. By studying neutrinos in plane symmetric spacetimes, physicists can isolate the effects of gravity on their trajectories, which can be obscured by matter in more complex spacetimes\cite{carlson-safko}. This isolation of gravity effects on neutrinos is especially relevant given the importance of gravitational fields in astrophysical objects.
The study of neutrinos in gravitational fields can test predictions of proposed extensions to the Standard Model, such as models that incorporate extra dimensions or modify the behavior of neutrinos at high energies.
The simplicity of plane symmetric spacetimes makes them an ideal environment for testing these predictions. In such spacetimes, the equations describing the behavior of neutrinos can be simplified, allowing for more precise theoretical predictions and easier comparison with experimental observations.

Einstein's theory of gravitation can be viewed naturally as a gauge theory of $SO(1,3)$, the structure group of local ortho-chronous Lorentz transformations.
In such approaches the gauge potentials are usually ascribed to an affine connection in the orthonormal frame bundle $OM$ over the spacetime manifold $M$ ( that is  equipped with a second rank covariant, symmetric non-degenerate metric tensor). The connection  in Einstein's theory is the unique metric compatible and torsion-free Levi-Civita connection. The local boson fields  are then assigned to various irreducible tensor representations of the local Lorentz group.
On the other hand, local fermion fields should be assigned to irreducible spinor representations of  the algebra $sl(2,{\mathcal{C}})$, the double cover of the local Lorentz group.

 Thus, a more basic starting point for a gauge approach to gravity involves a non-trivial $SL(2,{\mathcal{C}})$ connection in a bundle of spin frames over $M$ (See e.g.\cite{benn}). Then a linear connection that is metric compatible but need not be torsion-free is given in the associated bundle of $OM$.
The neutrino fields in Nature come in three-generations that are each electrically neutral and left-chiral. Anti-neutrinos are similarly right-chiral.
The chiral Weyl (anti-)neutrinos, unlike Majorana neutrinos, lack invariance under parity,
A consistent definition of spacetime frame reversals can be established by inducing on $OM$ discrete transformations defined essentially on the bundle of spin frames over $M$.
These carry linear or conjugate linear representations of operators  $P$,$T$ or $C$.The  identification of the components of spin-vectors (i.e.spinors) in these frames with the Lorentz group spinors imply  that such  intrinsic spacetime operations can be implemented by conventional spinor component transformations under the extended Lorentz group
$SL(2,{\mathcal{C}})$ to within an arbitrary set of phases.
Therefore the standard model  neutrinos   may be described by Weyl spinors transforming  according to the fundamental representation of  $SL(2,{\mathcal{C}})$ while the corresponding anti-neutrinos are described by Weyl spinors  that live in the conjugate representation.

It is a well established  fact that the algebra of complex quaternions forms a homomorphic image of the $sl(2,{\mathcal{C}})$ Lie algebra,thus enabling one to establish a natural
$SL(2,{\mathcal{C}})$ group action on tensor as well as spinor fields.
Since the irreducible representations of  $SL(2,{\mathcal{C}})$ play an essential role in the description of neutrinos and anti-neutrinos in Nature,  we find it convenient
to embed all the geometrical entities involved on $OM$ into complex quaternion algebra. This will afford freedom from, for instance,  Fierz rearrangements or other
tedious algebraic manipulations in the ring of $\gamma$-matrices involving the spinor fields. All these operations would be replaced by straightforward quaternion algebra rules.


 We use here the complex quaternion valued exterior differential forms over the 4-dimensional space-time \cite{dereli-tucker,benn}.
Exterior differential forms are coordinate independent and all relevant tensors in this language remain free of indices $\mu,\nu,\dots$ in a coordinate chart $\{x^{\mu}\}.$
The algebra of complex quaternions on the other hand carry representations of the spin cover of the local Lorentz group, that is, $SL(2,{\mathcal{C}})$. As such it can be used
to express  the local frames, connection and curvature forms without use of Lorentz indices $a,b,\dots$.  Thus an algebraic formulation of 4-dimensional space-time geometries is achieved without any indices. The massless chiral spinor fields are introduced in this language by considering minimal left ideals as spin basis\footnote{These correspond to undotted spinors in  the NP formalism. In our approach dotted spinors of NP formalism are identified as minimal right ideals with a corresponding spin basis.}.The advantage of our unconventional formalism is that once a local Lorentz frame is chosen, it automatically fixes the corresponding spin basis, so that one doesn't have to worry about determining
a set of convenient $\gamma$-matrices  that would blend in well with the chosen local co-frame.

\noindent The plan of the paper is the following.
In Section:2, complex quaternion valued exterior differential forms over 4-dimensional Lorentzian space-times are briefly introduced. Then we define Weyl spinor  fields
as minimal left ideals in the complex quaternion algebra. The metric and the connection with torsion are given and the corresponding Cartan structure equations are shown.
Section:3 is reserved for the variational derivation of the coupled Einstein-Weyl equations from an action. Both the first order variations where the co-frame and the linear connection are varied independently; and the second order variations where the connection is fixed to be the unique torsion-free Levi-Civita connection of the metric tensor are given.
We use the method of Lagrange multipliers for the torsion-free case.
The field equations for both first and second order variations are explicitly derived and simplified. It is pointed out that the equivalence of the field equations for the two cases can be insured by adjusting a particular quartic self-coupling of the Weyl spinor in the action.
We discuss exact plane symmetric solutions of the Einstein-Weyl equations in Section:4. We work in a complex null  co-frame
that are expressed in local coordinates adapted to the given complex null structure. Together with a generic Weyl spinor ansatz, the  reduced Einstein-Weyl equations
are worked out for both the zero-torsion and the non-zero torsion cases. They turned out to be the same, because of the simplifying  isometries of a plane symmetric space-time.
Finally we provide two simple families of exact solutions
 that describe either  left-moving or  right-moving neutrino plane waves.
 There are some brief concluding comments in Section:5.

\section{Complex Quaternionic Differential Forms}
All elements in the algebra of complex quaternions, denoted as $\mathbb{C}\otimes\mathbb{H}$, can be expressed as a linear combination of eight elements: $1$, $i$, $\hat{e_1}$, $\hat{e_2}$, $\hat{e_3}$, $i\hat{e_1}$, $i\hat{e_2}$, and $i\hat{e_3}$, where $\hat{e_1}$, $\hat{e_2}$, and $\hat{e_3}$ are quaternionic units in $\mathbb{H}$ that satisfy certain commutation relations
\begin{equation}
 \hat{e}_k\hat{e}_j=-\delta_{kj}+\epsilon_{kjl}\hat{e}_l.
\end{equation}
The complex unit $i$ commutes with every element in the algebra, and it follows that $i^2=-1$.
It is straightforward to associate the three objects $i\hat{e}_1$, $i\hat{e}_2$, and $i\hat{e}_3$ with the Pauli matrices $\sigma_1$, $\sigma_2$, and $\sigma_3$, which generate rotations. Similarly, the multiplication and commutation rules of $\hat{e}_1$, $\hat{e}_2$, and $\hat{e}_3$ demonstrate that they are responsible for boosts. These six items can be shown to produce the Lorentz algebra. To obtain an element $Q\in SL(2,\mathbb{C})$, one can take a real linear combination of the six items, multiply the result by complex $i$, and exponentiate the whole thing. Complex conjugation is defined by the map $(i \rightarrow -i)$ and denoted by a right-superscript $^*$, while quaternionic conjugation $(\hat{e}_k \rightarrow - \hat{e}_k)$ is indicated by an overbar. Their composition is denoted by $^\dagger$.

The elements $Q\in SL(2,{\mathbb{C}})$ are unit norm quaternions satisfying $Q\bar{Q}=\bar{Q}Q=1$. One way to obtain such an element is given by
$$
Q=e^{\frac{i}{2}\hat{a} \alpha} e^{\frac{1}{2}\hat{b} \beta}
$$
where $\hat{a}$ and $\hat{b}$ are unit q-vectors and $\alpha$ and $\beta$ are real parameters. Given a local Lorentz transformation $Q = q_4 + \sum q_k \hat{e}_k \in SL(2,\mathbb{C})$, one can introduce the matrix
$$
\tilde{Q} = \left (\begin{array}{cc} q_4 -iq_3 & -q_2 -iq_1 \\ q_2-iq_1 & q_4+iq_3 \end{array} \right )
$$
so that $Q\bar{Q}=\bar{Q}Q=1$ if and only if $det{\tilde{Q}}=1$. In what follows, we will use the same symbol to denote both the quaternion $Q$ and the matrix $\tilde{Q}$ without any potential confusion.

\smallskip

\noindent Consider a complex differential $p$-form $A$. We can express $A$ in terms of the basis elements $\hat{e}_k$ as follows:
\begin{equation}
A^{(p)}=A^{(p)}_4 + \sum_{k=1}^{3} A^{(p)}_k \hat{e}_k,
\end{equation}
where $A_4$, $A_1$, $A_2$, and $A_3$ are complex-valued $p$-forms.

\smallskip

\noindent We define the following operators acting on arbitrary quaternions $q$ in terms of complex and quaternionic conjugation:
\ba
2Re(q)=q+q^*, \quad 2iIm(q) = q - q^*, \nonumber \\  \quad 2Sc(q)=q+\bar{q}, \quad
2Vec(q)=q-\bar{q}, \nonumber \\ 2\mathcal{H}(q)=q+q^{\dagger}, \quad 2\mathcal{A}(q)=q-q^{\dagger}.
\ea
More precisely, we say that $q$ is a $q$-scalar when $Vec(q)=0$, a $q$-vector when $Sc(q)=0$, and Hermitian and anti-Hermitian when $\mathcal{H}(q)=q$ and $\mathcal{H}(q)=-q$, respectively.

\medskip

\noindent On the other hand, Weyl spinors of $SL(2,\mathbb{C})$ are represented by 2-component complex vectors that transform as follows:
\ba
\phi_{\alpha}^{\prime} = (Q\phi)_{\alpha} , \quad {\phi^{\alpha}}^{\prime} = (\phi \bar{Q})^{\alpha}
\ea
for undotted spinors and
\ba
\psi_{\dot{\alpha}}^{\prime} = (\psi Q^{\dagger})_{\dot{\alpha}} , \quad {\psi^{\dot{\alpha}}}^{\prime} = (Q^{*}\psi)^{\dot{\alpha}}
\ea
for dotted spinors.
These Weyl spinors can be represented in the algebra of complex quaternions by left ideals generated by $L_1:(U^1,U^2)$ and $L_2:(W^1,W^2)$, or the right ideals generated by $R_1:(U^1,W^2)$ and $R_2:(U^2,W^1)$, where
\ba
U^1=\frac{1}{\sqrt{2}}(1+i\hat{e}_3), \quad U^2=\frac{1}{\sqrt{2}}(\hat{e}_2+i\hat{e}_1), \nonumber \\
W^1=\frac{1}{\sqrt{2}}(1-i\hat{e}_3), \quad W^2=\frac{1}{\sqrt{2}}(\hat{e}_2-i\hat{e}_1).
\ea
The multiplication table for the elements of these spin bases is given below. (An entry in the first row is  multiplied from the left by an entry in the first column.)

\medskip

\begin{center}
\begin{tabular}{l|cccc}
& $U^1$ & $U^2$ & $W^1$ & $W^2$ \\
\hline
 $U^1$ &  $\sqrt{2}U^1$ & $0$   &  $0$  & $\sqrt{2}W^2$ \\
 $U^2$ &  $\sqrt{2}U^2$  &  $0$  &  $0$  & $-\sqrt{2}W^1$  \\
 $W^1$ &   $0$   & $\sqrt{2}U^2$   &  $\sqrt{2}W^1$  &  $0$ \\
 $W^2$ &    $0$  &$-\sqrt{2}U^1$    & $\sqrt{2}W^2$   & $0$ \\
\end{tabular}
\end{center}

\bigskip

\noindent The left and right action of a unit quaternion $Q$ ($Q\bar{Q}=\bar{Q}Q=1$) can induce an $SL(2,\mathbb{C})$ transformation on spinors, as demonstrated by the equations given below:
\begin{alignat}{4}
\phi&=\phi_1U^1+\phi_2U^2 &&\rightarrow&& \hspace{2mm} Q\phi , \nonumber \\
\dot{\chi}&=\chi_{\dot{1}}W^1+\chi_{\dot{2}}W^2 &&\rightarrow&& \hspace{2mm} Q^{*}\dot{\chi} , \nonumber \\
\psi&=\psi_{1}U^1+\psi_{2}W^2 &&\rightarrow&& \hspace{2mm} \psi \bar{Q} , \nonumber \\
\dot{\xi}&=\xi_{\dot{1}}U^2+\xi_{\dot{2}}W^1 &&\rightarrow&& \hspace{2mm} \dot{\xi} Q^{\dagger}.
\end{alignat}

\medskip

\noindent A Majorana spinor valued p-form can be expressed in terms of two independent complex spinor p-forms, denoted by $\phi_A$ and $\phi_B$. Then
\begin{eqnarray}
\phi_{+} &=& -\phi_B U^1 + \phi_{A} U^2, \quad  \phi_{-} = \phi_A W^1 + \phi_{B} W^2, \nonumber \\ {\dot{\phi}}_{+} &=&  i \left ( \phi_B^{*} U^1 + \phi_{A}^{*} W^2 \right ),
\quad  {\dot{\phi}}_{-} = i \left ( \phi_{A}^{*}  W^1-\phi_{B}^{*} U^2 \right ).
\end{eqnarray}
In general, Majorana spinor valued p-forms satisfy
\ba
{\phi^{\dagger}_{\pm}}= \pm i {\dot{\phi}}_{\pm} .
\ea

\medskip

\noindent Below is a brief explanation of the 4-dimensional space-time geometry, presented in the language of complex quaternion valued exterior differential forms. Firstly, we examine the (anti-Hermitian) co-frame 1-form, which can be written as follows:

\begin{equation}
e=ie^0+\sum_{k=1}^{3}e^k\hat{e}_k=-e^{\dagger}
\end{equation}
The local coordinate chart ${x^{\mu}}$ defines the usual tetrad components ${e^{a}_{\mu}}$ as $e^a=e^a_{\mu}dx^{\mu}$, where $a=0,1,2,3$. An $SL(2,\mathbb{C})$ transformation can be applied to the co-frame $e$ according to
\ba
e \rightarrow QeQ^{\dagger}.
\ea
Using this basis 1-form, we can represent the metric of spacetime in the following way:
\ba
g=Re(e\otimes\bar{e}),
\ea
We define the connection, torsion, and curvature forms over space-time as follows:
\ba
\omega&=&\sum_{k=1}^{3}\omega^k\hat{e_k}, \nonumber \\
T&=&de + \omega \w e - e \w \omega^{\dagger}=iT^0+\sum_{k=1}^{3}T^k\hat{e_k}=-T^{\dagger}, \nonumber \\
R&=&d\omega+\omega \w \omega = \sum_{k=1}^{3}R^k\hat{e_k},
\ea
with
\ba
\omega^k=-\frac{1}{2}(i\omega^0_{\;\;k}+\frac{1}{2}\epsilon_{ijk}\omega^{i}_{\;\;j}), \quad
R^k=-\frac{1}{2}(iR^0_{\;\;k}+\frac{1}{2}\epsilon_{ijk}R^{i}_{\;\;j})
\ea
where $\omega_{ab}=-\omega_{ba}$ are the real components of the connection 1-forms and where $i,j,k, . . . = 1,2,3$ and the symbol $\epsilon_{ijk}$ is totally antisymmetric with $\epsilon_{123} = +1$.

The connection 1-forms $\omega_{ab}$ have real components, and the frame indices are raised and lowered through the Minkowski metric $\eta_{ab}=diag(-,+,+,+)$. The torsion $T$ is anti-Hermitian, so its components $(T^0,T^1,T^2,T^3)$ are real 2-forms. The connection and curvature forms $\omega$ and $R$ are $SL(2,\mathbb{C})$ valued, and thus, $\omega^k$ and $R^k$ are complex 1- and 2-forms, respectively.
Using the aforementioned structural equations, which serve as their integrability requirements, we may deduce the following Bianchi identities:
\ba
dT + \omega \wedge T -T \wedge \omega^{\dagger} = R \wedge e + e \wedge R^{\dagger},
\ea
and
\ba
dR + \omega \wedge R - R \wedge \omega = 0.
\ea
It is noteworthy that the verification of the following local Lorentz transformation rules is straightforward:
$$
\omega \rightarrow Q \omega \bar{Q} + Qd\bar{Q}, \quad  T \rightarrow O T Q^{\dagger}, \quad R \rightarrow Q R \bar{Q}.
$$
Consider a Weyl spinor that transforms as $\phi \rightarrow Q\phi$. Then we define its  covariant exterior derivative
$$
\nabla \phi = d\phi + \omega \phi
$$
which transforms properly under a local Lorentz transformation as $\nabla \phi \rightarrow  Q \nabla \phi$.  Similarly for a dotted spinor we have $\dot{\chi} \rightarrow Q^{*}\dot{\chi}$ so that
$$
\nabla \dot{\chi} = d \dot{\chi} -  \omega^{\dagger}\dot{\chi}.
$$

\medskip

\noindent Given that we will be utilizing complex null basis 1-forms often in our calculations, we will now go over their fundamentals. First and foremost, the connections between the complex null co-frames and the orthonormal co-frames are fixed as
\ba
l=\frac{1}{\sqrt{2}} (e^3+e^0), \quad n=\frac{1}{\sqrt{2}} (e^3-e^0), \quad m=\frac{1}{\sqrt{2}}(e^1+ie^2 ).
\ea
The metric becomes
\begin{equation}
  g=l \otimes n +n \otimes l  + m \otimes m^{*} + m^{*} \otimes m;
\end{equation}
and the orientation of spacetime is defined by $*1=e^0 \w e^1 \w e^2 \w e^3 =i l \w n \w m \w m^{*}$, where $*$ to the left stands for the Hodge dual defined on forms.
On the other hand, one should be aware that a $*$ to the right of a sign implies complex conjugation.
At this point, we make transition to a more practical notation by setting
\ba
\omega_{\pm}=\omega^1\pm i\omega^2,\quad \omega_0=\omega^3, \quad
R_{\pm}=R^1\pm iR^2, \quad  R_0=R^3.
\ea
Now, these may be expressed directly in terms of the spinor basis elements and null tetrads as follows\ba
e&=&iW^1l -iU^1n-iU^2m+iW^2m^{*}, \nonumber \\
\omega &=&\frac{i}{\sqrt{2}} \omega_{-} \; W^2 -\frac{i}{\sqrt{2}} \omega_{0} \; (U^1-W^1) -\frac{i}{\sqrt{2}} \omega_{+} \; U^2 ,  \nonumber \\
R &=&\frac{i}{\sqrt{2}} R_{-} \; W^2- \frac{i}{\sqrt{2}}R_{0}\;(U^1-W^1) -\frac{i}{\sqrt{2}}R_{+} \; U^2.
\ea

\section{Einstein-Weyl Theory}

\noindent  We discuss now the minimal coupling of a Weyl neutrino field to Einstein's gravity. Since the neutrino is described in conventional quantum field theory by a chiral spinor field in the fundamental representation of the extended Lorentz group, we are immediately faced with two essential interpretational problems.
Firstly, the spin-statistics relation requires the spinor components to be mutually anti-commuting. Secondly, we face the problem of making sense of  quadratic neutrino expressions in the stress-energy-momentum tensor in the Einstein field equations that may also involve quartic neutrino self-couplings, which are  both essentially terms of quantum nature.
For the purpose of generating dynamical field equations by a variational principle from an action, we treat components of Fermi spinors as mutually anti-commuting (hence,nilpotent) odd-elements of an underlying abstract Grassmann algebra. All the manipulations leading to exact solutions below will be done under this assumption.
However, such solutions remain for further examination whether they provide sensible models in terms of quantum operator algebras to describe  neutrino-gravity interactions.
This will not be attempted here.

The coupled system of Einstein-Weyl field equations are determined by a variational principle from an action
\ba
I[e,\omega,\xi] = \int_{M} \mathcal{L}
\ea
where the Lagrangian density 4-form in terms of complex quaternionic exterior differential forms is given by
\ba
\mathcal{L}=ImSc(2R\w e \w e^* + 2i ^*\bar{e} \w \nabla \xi \xi^{\dagger}).
\ea
We will vary the action with respect to independent variables \{$ e,\omega,\xi$\}.
In a first order formulation, the co-frame $e$ is varied rather than the metric. Furthermore the connection carries torsion and hence it can be varied independently of the metric.
On the other hand in a second order formulation the connection is fixed to be the Levi-Civita connection of the metric and therefore can't be varied independently. We  deal with this problem here by the method of Lagrange multipliers. We introduce a set of Lagrange multiplier 2-forms $\lambda_a$ that will be varied freely. The fact that torsion of space-time vanishes is imposed as a constraint.  Then the variations of the action with respect to  \{$ e,\omega,\xi, \lambda$\} impose the zero-torsion constraint. The remaining field equations are solved subject to this constraint. The field equations implied by the connection variations can then be solved for the multipliers which are substituted in to the remaining coupled Einstein-Weyl equations.

\subsection{First order variational field equations}

\noindent

\noindent We first consider the co-frame variations of $\mathcal{L}$. We will be using the identity $^*\bar{e}=\frac{i}{6}e^* \w e \w e^*$. Then\footnote{Here and all that follow, variations are given
up to  closed forms.}
\ba
\delta_e\mathcal{L} &=&ImSc\Big(2R \w \delta e \w e^* + 2R \w e \w \delta e^* - \frac{1}{3}\delta e^* \w e \w e^* \w \nabla\xi\xi^{\dagger} \nonumber \\
 &-& \frac{1}{3} e^* \w \delta e \w e^* \w \nabla\xi\xi^{\dagger} - \frac{1}{3} e^* \w e \w \delta e^* \w \nabla\xi\xi^{\dagger} \Big) \nonumber \\
 &=&ImSc\Big(-2e^* \w R \w \delta e + 2R \w e \w \delta e^* + \frac{1}{3}e \w e^* \w \nabla\xi\xi^{\dagger}\w \delta e^*\nonumber \\
 &-& \frac{1}{3} e^* \w \nabla\xi\xi^{\dagger} \w e^* \w \delta e + \frac{1}{3} \nabla\xi\xi^{\dagger} \w e^* \w e \w \delta e^*\Big) ,\nonumber \\
\ea
since we can rearrange quaternionic forms under the scalar sign.
Since $Im(X)=-Im(X^*)$, we have
\ba
\delta_e\mathcal{L} &=&ImSc\Big(2e^* \w R \w \delta e^*+ 2R \w e \w \delta e^* + \frac{1}{3}e \w e^* \w \nabla\xi\xi^{\dagger}\w \delta e^*\nonumber \\
&+& \frac{1}{3} e\w \nabla\xi\xi^{\dagger} \w e \w \delta e^*+ \frac{1}{3} \nabla\xi\xi^{\dagger}\w e^* \w e \w  \delta e^*\Big) \nonumber \\
&=&ImSc\Big(\big(2e^* \w R+ 2R \w e + \frac{1}{3}e \w e^* \w \nabla\xi\xi^{\dagger} + \frac{1}{3} e \w \nabla\xi\xi^{\dagger}\w e  \nonumber \\
&+&\frac{1}{3} \nabla\xi\xi^{\dagger} \w e^* \w e ) \w \delta e^{*} \Big).
\ea
One may as well write
\ba
\delta_e\mathcal{L}&=&2ImSc\Big(\mathcal{H}\big(2R \w e + \frac{1}{6}(e \w e^* \w \nabla\xi\xi^{\dagger}+e \w \nabla\xi\xi^{\dagger}\w e+\nabla\xi\xi^{\dagger} \w e^* \w e))\w \delta e^{*}  \Big), \nonumber
\ea
so that the Einstein field equation becomes
\ba
-2\mathcal{H}(R \w e)=\frac{1}{6}(e \w e^* \w \nabla\xi\xi^{\dagger}+e \w \nabla\xi\xi^{\dagger}\w e+\nabla\xi\xi^{\dagger} \w e^* \w e) .
\ea
The term  on the right hand side  is identified as the canonical stress-energy-momentum 3-form of a Weyl neutrino. It is asymmetrical in general.

\medskip

\noindent Next the connection variations of the action density 4-form give
\ba
\delta_{\omega}\mathcal{L}&=&ImSc\Big( 2(d\delta \omega + \delta \omega \w \omega + \omega \w \delta \omega) \w e \w e^* - \frac{1}{3}e^* \w e \w e^* \delta \omega \xi\xi^{\dagger}\Big) \nonumber \\
&=& ImSc\Big( \delta \omega \w 2(d(e\w e^*)+\omega \w e \w e^* + e \w e^* \omega) + \frac{1}{3}  \delta \omega \w \xi\xi^{\dagger} e^* \w e \w e^* \Big),\nonumber \\
&=&ImSc\Big( \delta \omega \w \big(2\nabla(e\w e^*) + \frac{1}{3} \xi\xi^{\dagger} e^* \w e \w e^*\big)  \Big).
\ea
Then the equations to be solved read
\ba
ImSc\Big( \delta \omega \w \big(2\nabla(e\w e^*) + \frac{1}{3} \xi\xi^{\dagger} e^* \w e \w e^*\big)  \Big)=0.
\ea
Since $\omega$ is a complex $q$-vector valued 1-form, we have
\ba
Vec[2T\w e^*+\frac{1}{6}\xi\xi^{\dagger}e^*\w e\w e^*]=0.
\ea
Suppose $h=\frac{1}{6}\xi\xi^{\dagger}$ is a Hermitian 0-form. Then the above field equation becomes
\ba
Vec \left (2T \w e^* + h e^* \w e \w e^*\right ) =0.
\ea
Because $Vec(Sc(X))=0$ by definition, we write
\ba
&&Vec[2T \w e^* + h e^* \w e \w e^*+2Sc(e \w e^* \w he^*)]=0, \nonumber \\
&&Vec[2T \w e^* + h e^* \w e \w e^*+e \w e^* \w he^*+e\w \bar{h}e \w e^*]=0, \nonumber \\
&&Vec[(2T + 2\mathcal{A}(he^*\w e)+e\w \bar{h}e) \w e^*]=0.
\ea
We note that the above complex $q$-vector valued 3-form equation can be broken apart as twenty four algebraic equations for  twenty four components of an anti-Hermitian 2-form. Thus it admits  the unique solution
\ba
2T=-2\mathcal{A}(he^*\w e)-e\w \bar{h}e.\label{torsion}
\ea
Therefore the field equation resulting from the connection variations determines the space-time torsion algebraically by a quadratic expression in terms of a Weyl spinor $\xi$.

\medskip

\noindent Finally  we determine the variations of $\mathcal{L}$ with respect to the spinor field:
\ba
\delta_{\xi}\mathcal{L}&=&ImSc\Big(2i^*\bar{e}\w \nabla(\delta \xi) \xi^{\dagger}+2i*\bar{e}\w \nabla\xi \delta\xi^{\dagger}\Big),\nonumber \\
&=&ImSc\Big(2i^*\bar{e}\w \nabla\xi \delta\xi^{\dagger}+2i\nabla(^*\bar{e})\delta \xi \xi^{\dagger}-2i^*\bar{e}\w \delta \xi \nabla\xi{\dagger}-2i\nabla(^*\bar{e}\delta \xi \xi^{\dagger})\Big) \nonumber \\
&=&ImSc\Big(2i^*\bar{e}\w \nabla\xi \delta\xi^{\dagger} - 2i\nabla(^*\bar{e})\xi \delta \xi^{\dagger}+ 2i^*\bar{e}\w \nabla\xi \delta\xi^{\dagger} \Big) \nonumber \\ &=&ImSc\Big((4i^*\bar{e}\w \nabla\xi  - 2i\nabla(^*\bar{e})\xi)\w \delta\xi^{\dagger} \Big) .
\ea
Then since $ImSc(X)=-ImSc(X^{\dagger})$,
the field equation becomes
\ba
^*\bar{e}\w \nabla\xi - \frac{1}{2}(\nabla^*\bar{e})\xi=0.
\ea
Apparently the Weyl equation above has picked up an explicit
non-linear self-coupling term. However, this is not the case. Because of the identity  $^*\bar{e}=\frac{1}{6}e^* \w e \w e^*$, we have
\ba
\nabla^*\bar{e} &=&\frac{i}{6}(\nabla e^* \w e \w e^* - e^* \w \nabla e \w e^* + e^* \w e \w \nabla e^*) \nonumber \\
&=&\frac{i}{6}(T^*\w e \w e^*- e^* \w T \w e^* +e^* \w e \w T^*)\nonumber \\
&=&\frac{i}{12}(e^* \w eh^* \w e \w e^*) = 0.
\ea
A detailed derivation is given in the Appendix.
Therefore the Weyl  equation simply reads
\ba
^*\bar{e}\w \nabla\xi=0.
\ea
To summarize,  we end up with the system of coupled Einstein-Weyl field equations (25),(31) and (35).

\subsection{Second order variational field equations}

\noindent In terms of complex quaternionic exterior differential forms, now we start with the Lagrangian  density 4-form
\ba
{\mathcal{L}}^{\prime}=ImSc(2R\w e \w e^* + 2i ^*\bar{e} \w \nabla\xi \xi^{\dagger} + i\bar{\lambda}\w T)
\ea
where $\lambda=i\lambda^0+\lambda^k\hat{e_k}$ are related with the Lagrange multiplier 2-forms \{$\lambda_a$\}. We remind the definition of  the torsion 2-form:
\ba
T=de+2\mathcal{A}(\omega \w e) = de + \omega \w e - e \w \omega^{\dagger}.
\ea
Let's consider the infinitesimal variations of ${\mathcal{L}}^{\prime}$ with respect to  independent variables \{$ e,\omega,\lambda,\xi$\}.
From the co-frame variations, using the identity $^*\bar{e}=\frac{i}{6}e^* \w e \w e^*$, we get
\ba
\delta_e{\mathcal{L}}^{\prime}&=&ImSc\Big(2R \w \delta e \w e^* + 2R \w e \w \delta e^* - \frac{1}{3}\delta e^* \w e \w e^* \w \nabla\xi\xi^{\dagger} \nonumber \\
 &-& \frac{1}{3} e^* \w \delta e \w e^* \w \nabla\xi\xi^{\dagger} - \frac{1}{3} e^* \w e \w \delta e^* \w \nabla\xi\xi^{\dagger}\nonumber \\
 &+&i\bar{\lambda}\w(d\delta e + \omega \w \delta e - \delta e \w \omega^{\dagger}) \Big) \nonumber \\
 &=&ImSc\Big(-2e^* \w R \w \delta e + 2R \w e \w \delta e^* + \frac{1}{3}e \w e^* \w \nabla\xi\xi^{\dagger}\w \delta e^*\nonumber \\
 &-& \frac{1}{3} e^* \w \nabla\xi\xi^{\dagger} \w e^* \w \delta e + \frac{1}{3} \nabla\xi\xi^{\dagger} \w e^* \w e \w \delta e^* \nonumber \\
 &+&i(-d\bar{\lambda}\w \delta e +\bar{\lambda} \w \omega \w \delta e + \omega^{\dagger}\w \bar{\lambda} \w  \delta e) \Big),
\ea
because quaternionic forms can be re-arranged under the scalar sign.
Since $Im(X)=-Im(X^*)$, we have
\ba
\delta_e{\mathcal{L}}^{\prime}&=&ImSc\Big(2e^* \w R \w \delta e^*+ 2R \w e \w \delta e^* + \frac{1}{3}e \w e^* \w \nabla\xi\xi^{\dagger}\w \delta e^*\nonumber \\&+& \frac{1}{3} e\w \nabla\xi\xi^{\dagger} \w e \w \delta e^*+ \frac{1}{3} \nabla\xi\xi^{\dagger}\w e^* \w e \w  \delta e^* \nonumber \\  &+&i(-d\lambda^{\dagger} +\lambda^{\dagger} \w \omega^* +\bar{\omega} \w \lambda^{\dagger}) \w \delta e^* \Big) \nonumber \\
&=&ImSc\Big(\big(2e^* \w R+ 2R \w e + \frac{1}{3}e \w e^* \w \nabla\xi\xi^{\dagger} + \frac{1}{3} e \w \nabla\xi\xi^{\dagger}\w e  \nonumber \\
&+&\frac{1}{3} \nabla\xi\xi^{\dagger} \w e^* \w e + i\nabla\lambda \big) \w \delta e^* \Big)
\ea
where we used $\lambda^{\dagger}=-\lambda$ and $\bar{\omega}=-\omega$.
Hence
\ba
\delta_e {\mathcal{L}}^{\prime}&=&2ImSc\Big(\mathcal{H}\big(2R \w e + \frac{1}{6}(e \w e^* \w \nabla\xi\xi^{\dagger}+e \w \nabla\xi\xi^{\dagger}\w e+\nabla\xi\xi^{\dagger} \w e^* \w e \nonumber \\
&+&\frac{i}{2}\nabla\lambda\big) \w \delta e^* \Big),
\ea
and the Einstein field equation reads
\ba
-2\mathcal{H}(R \w e)=\frac{1}{6}(e \w e^* \w \nabla\xi\xi^{\dagger}+e \w \nabla\xi\xi^{\dagger}\w e+\nabla\xi\xi^{\dagger} \w e^* \w e) + \frac{i}{2}\nabla\lambda.
\ea

\noindent Next the connection variations of the action density 4-form give
\ba
\delta_{\omega}{\mathcal{L}}^{\prime}&=&ImSc\Big( 2(d\delta \omega + \delta \omega \w \omega + \omega \w \delta \omega) \w e \w e^* - \frac{1}{3}e^* \w e \w e^* \delta \omega \xi\xi^{\dagger} \nonumber \\
&+&i\bar{\lambda}\w( \delta \omega \w e - e \w \delta \omega^{\dagger}) \Big) \nonumber \\
&=& ImSc\Big( \delta \omega \w 2(d(e\w e^*)+\omega \w e \w e^* + e \w e^* \omega) + \frac{1}{3}  \delta \omega \w \xi\xi^{\dagger} e^* \w e \w e^* \nonumber \\
&+&i\delta \omega \w e \w \bar{\lambda} + i\delta \omega^{\dagger} \w \bar{\lambda} \w e ) \Big),\nonumber \\
&=&ImSc\Big( \delta \omega \w \big(2 \nabla(e\w e^*) + ie\w \bar{\lambda}-i\lambda^{\dagger} \w e^* + \frac{1}{3} \xi\xi^{\dagger} e^* \w e \w e^*\big)  \Big), \nonumber \\
&=&ImSc\Big( \delta \omega \w \big(2\nabla(e\w e^*) + i(e\w \bar{\lambda}+\lambda \w e^*) + \frac{1}{3} \xi\xi^{\dagger} e^* \w e \w e^*\big)  \Big). \nonumber\\
\ea
We notice that the $\delta \lambda$ variation of the action density 4-form imposes the zero torsion constraint $T=0$, which in turn implies that the term $\hat{\nabla}(e\w e^*)$ above vanishes identically. $\hat{\nabla}$ denotes the Levi-Civita connection,
so the equation to be solved algebraically for $\lambda$ becomes
\ba
e\w \bar{\lambda}+\lambda \w e^*=\frac{i}{3} \xi\xi^{\dagger} e^* \w e \w e^*.
\ea

\noindent The variation with respect to the spinor field is identical to that in the previous section, plus the fact that the torsion vanishes. Thus we get the Weyl equation
\ba
^*\bar{e}\w \hat{\nabla}\xi=0,
\ea
that is coupled to the Einstein field equation
\ba
-2\mathcal{H}(\hat{R} \w e)=\frac{1}{6}(e \w e^* \w \hat{\nabla}\xi\xi^{\dagger}+e \w \hat{\nabla}\xi\xi^{\dagger}\w e+\hat{\nabla}\xi\xi^{\dagger} \w e^* \w e) + \frac{i}{2}\hat{\nabla}\lambda
\ea
where one should still substitute in $\lambda$  from above.
We note that the sum of the two terms on the right hand side of the Einstein equation could be identified as the symmetrical stress-energy-momentum 3-form of a Weyl neutrino.

\section{Exact Plane Symmetric Solutions}

\noindent The basic field variables of the theory consists of a co-frame field $e$ in terms of which the Lorentzian metric
\ba
g= Re(e \otimes \bar{e})
\ea
and a Weyl spinor field $\xi$ that is odd-Grassmann valued.
 Let us start with a plane symmetric spacetime metric
\begin{equation}
  g = e^{2\mu(u,v)} 2 du \otimes dv+e^{2\nu(u,v)}2 d\zeta \otimes d\zeta^*,
\end{equation}
given in terms of null coordinates
\ba
u = \frac{z+t}{\sqrt{2}}, \quad v = \frac{z-t}{\sqrt{2}},\quad \zeta=\frac{x+iy}{\sqrt{2}}, \quad \zeta^{*} = \frac{x-iy}{\sqrt{2}} .
\ea
Next we introduce the complex null basis 1-forms
\ba
l=e^{\mu} du, \quad n = e^{\mu} dv, \quad  m= e^{\nu} d\zeta, \quad m^{*} = e^{\nu} d\zeta^{*} .
\ea
The corresponding Levi-Civita connection 1-forms in the complex null basis turn out to be
\ba
{\hat{\omega}}_{+} &=& i e^{-\mu}(\frac{\partial \nu}{\partial v}) m , \nonumber \\
{\hat{\omega}}_{0} &=& -\frac{i}{2}  e^{-\mu} (\frac{\partial \mu}{\partial u})   l +\frac{i}{2}  e^{-\mu} (\frac{\partial \mu}{\partial v})  n , \nonumber \\
{\hat{\omega}}_{-} &=&-i  e^{-\mu} (\frac{\partial \nu}{\partial u}) m^{*} .
\ea

\noindent Let the neutrino field  be given by the ansatz
\ba
\xi = \xi_{1}(u,v,\zeta,\zeta^*) U^1 + \xi_{2}(u,v,\zeta,\zeta^*) U^2,
\ea
where $\xi_{1}$ and $\xi_{2}$ are arbitrary (odd-Grassmann valued) complex functions of all coordinates.



\subsection{ The case of zero-torsion:}

\noindent We will be looking for exact solutions of the Einstein field equation
\ba
-2\mathcal{H}(\hat{R} \w e)=\frac{1}{6}(e \w e^* \w \mathcal{A}(q)+e \w \mathcal{A}(q^*)\w e+\mathcal{A}(q) \w e^* \w e) + \frac{i}{2}\hat{D}\lambda.
\ea
where $q=\hat{\nabla}\xi\xi^{\dagger}$  coupled with the Weyl equation
\ba
{}^{*}\bar{e} \wedge \hat{\nabla} \xi = 0 .
\ea
The Weyl equation decouples in general in a complex null basis as
\begin{eqnarray}
l \wedge m \wedge m^{*} \wedge (  d\xi_1 -i{\hat{\omega}}_{0} \xi_1 -i{\hat{\omega}}_{-} \xi_2 ) - l \wedge n \wedge m^{*} \wedge ( d\xi_2 +i{\hat{\omega}}_{0} \xi_2 -i{\hat{\omega}}_{+} \xi_1) &=& 0 , \nonumber \\
n \wedge m \wedge m^{*} \wedge (  d\xi_2 +i{\hat{\omega}}_{0} \xi_2 -i{\hat{\omega}}_{+} \xi_1   ) + l \wedge n \wedge m \wedge (  d\xi_1 -i{\hat{\omega}}_{0} \xi_1 -i{\hat{\omega}}_{-} \xi_2 ) ) &=& 0 \nonumber .
\end{eqnarray}
We substitute in the corresponding  expressions in plane symmetric space-times  and after simplifications obtain the following set of coupled first order equations:
\ba
e^{-\mu} \left ( \frac{\partial \xi_1}{\partial v} +( \frac{\partial \nu}{\partial v} +\frac{1}{2} \frac{\partial \mu}{\partial v}) \xi_1 \right )  + e^{-\nu} \frac{\partial \xi_2}{\partial \zeta} &=&0, \\
-e^{-\nu} \frac{\partial \xi_1}{\partial  \zeta^*} + e^{-\mu} \left ( \frac{\partial \xi_2}{\partial u} + (\frac{\partial \nu}{\partial u}+\frac{1}{2} \frac{\partial \mu}{\partial u}) \xi_2 \right )&=&0.
\ea
Thus the general solution of the Weyl equation in plane symmetric spacetimes will be given by
\ba
\xi_1 = h_{1}(u,\zeta) e^{-(\nu + \frac{\mu}{2})}, \quad \xi_2= h_{2}(v,\zeta^*) e^{-(\nu + \frac{\mu}{2})} .
\ea

\noindent Next we work out the  Einstein field equations. They amount to the following set of  coupled differential equations:
\ba
2i \left ( \frac{\partial^2 \nu}{\partial u^2}+(\frac{\partial \nu}{\partial u})^2-2 (\frac{\partial \mu}{\partial u}) (\frac{\partial \nu}{\partial u}) \right )
&=&\frac{e^{-2 \nu}}{\sqrt{2}} \left ( \frac{\partial h_1}{\partial u} {h_1}^* - h_1 \frac{\partial {h_{1}}^*}{\partial u} \right ), \nonumber \\
-2i \left ( \frac{\partial^2 \nu}{\partial v^2}+(\frac{\partial \nu}{\partial v})^2-2 (\frac{\partial \mu}{\partial v}) (\frac{\partial \nu}{\partial v}) \right )
&=&\frac{e^{-2 \nu}}{\sqrt{2}} \left ( \frac{\partial h_2}{\partial v} {h_2}^* - h_2 \frac{\partial {h_{2}}^*}{\partial v} \right ), \nonumber \\
 \frac{\partial^2 \nu}{\partial u \partial v}+ 2 (\frac{\partial \nu}{\partial u}) (\frac{\partial \nu}{\partial v}) &=&0, \nonumber \\
 \frac{\partial^2 \mu}{\partial u \partial v}+\frac{\partial^2 \nu}{\partial u \partial v} + (\frac{\partial \nu}{\partial u}) (\frac{\partial \nu}{\partial v}) &=& 0, \nonumber \\
e^{-\nu+\mu}\frac{\partial h_1}{\partial \zeta} h_1^*+\left ( \frac{\partial h_1}{\partial u}+h_1(\frac{\partial \nu}{\partial u}-\frac{\partial \mu}{\partial u}) \right )h_2^{*} &=&0, \nonumber \\
e^{-\nu+\mu}\frac{\partial h_2}{\partial \zeta^{*}}  h_2^*-\left ( \frac{\partial h_2}{\partial v}+h_2 (\frac{\partial \nu}{\partial v}-\frac{\partial \mu}{\partial v}) \right )h_1^{*} &=&0, \nonumber \\
h_{2} (\frac{\partial h_1}{\partial \zeta})^{*} - (\frac{\partial h_2}{\partial \zeta^{*}}) h_1^* &=& 0.
\ea

\medskip

\subsection{The case of non-zero-torsion}

\noindent Using the torsion equation (\ref{torsion}), we calculate the following contortion 1-forms
\ba
K_{+} &=& -\frac{1}{2\sqrt{2}}(h_2h_1^*l+|h_2|^2m)e^{-2\nu-\mu},\nonumber \\
K_{0}&=& \frac{1}{4\sqrt{2}}(h_2h_1^*m^*-h_1h_2^*m-|h_1|^2l-|h_2|^2n)e^{-2\nu-\mu},\nonumber \\
K_{-} &=& \frac{1}{2\sqrt{2}}(h_1h_2^*n-|h_1|^2m^*)e^{-2\nu-\mu}.
\ea
The action of the contortion 1-forms on the neutrino equation is zero because of the odd-Grassman nature of the $h_1(u,\zeta)$ and $h_2(v,\zeta^*)$ and the remaining terms cancel each other.
The affine connection 1-form $\omega$ can be uniquely decomposed in to two parts according to
\ba
\omega = \hat{\omega} + K
\ea
where the Levi-Civita connection 1-form $\hat{\omega}$ satisfies the torsion-free Cartan structure equation
$$
d\hat{\omega} + \hat{\omega} \wedge e -e \wedge \hat{\omega}^{\dagger} = 0.
$$
The contortion 1-form $K$ is related to the torsion 2-form $T$ algebraically by
$$
K \wedge e -e \wedge K^{\dagger} = T.
$$
Thus the full curvature 2-form $R$ can also be decomposed according to
\ba
R = \hat{R} + \hat{\nabla}K + K \wedge K
\ea
where $\hat{\nabla} K = dK + \hat{\omega} \wedge K + K \wedge \hat{\omega}$ denotes the covariant exterior derivative relative to the Levi-Civita connection.
The coupled Einstein-Weyl equations then read
\begin{eqnarray}
-2{\cal{H}}(\hat{R} \wedge e) &=& 2{\cal{H}}(\hat{\nabla} K) + 2{\cal{H}}( K \wedge K \wedge e)   \nonumber \\
&& +\frac{1}{6} \left ( e \wedge e^{*} \wedge \hat{\nabla}\xi \xi^{\dagger} + e \wedge \hat{\nabla}\xi \xi^{\dagger} \wedge e + \hat{\nabla}\xi \xi^{\dagger} \wedge e^{*} \wedge e  \right ) \nonumber \\
&& +\frac{1}{6} \left ( e \wedge e^{*} \wedge K \xi \xi^{\dagger} + e \wedge K \xi \xi^{\dagger} \wedge e + K \xi \xi^{\dagger} \wedge e^{*} \wedge e  \right ), \nonumber
\end{eqnarray}
and
\ba
{}^{*}\bar{e} \wedge \hat{\nabla}\xi +   {}^{*}\bar{e} \wedge K \xi = 0.
\ea
In the Einstein field equations, the terms on the left hand side coming from the contortion part of the connection are carried to the right hand side and adds to the canonical energy-momentum tensor. Then we observe that exactly the same field equations as in the torsion-free case are obtained, that is, the quartic terms from the left hand side and the right hand side cancel each other out and  thus the remaining expression for the energy-momentum tensor is symmetrical.
Finally, the reduced field equations with non-zero torsion turns out to be exactly equal to those that are given above for the case of zero-torsion  in plane symmetric geometries.

\subsection{Plane wave solutions}

\noindent We discuss only two simple cases that describe either
\newline
(i) left-moving progressive waves where
\ba
\nu(u), \; \mu(u), \; h_1(u) \neq 0,  \; h_2=0
\ea
or
 \newline
 (ii) right-moving progressive waves where
\ba
\nu(v), \; \mu(v), \; h_1 = 0, \; h_2(v) \neq 0.
\ea
 In both these cases the Einstein-Weyl equations are satisfied provided the metric functions and the neutrino functions are related by the ordinary differential equation
 \ba
 \nu^{\prime \prime} +{\nu^{\prime}}^2 - 2 \mu^{\prime} \nu^{\prime} =  \mp \frac{i}{2\sqrt{2}} ( h^{\prime}h^{*} - h h^{* \prime} )
 \ea
 where ${}^{\prime}$  denotes differentiation by either $u$ or $v$, as the case maybe.  One may introduce a polar decomposition
 $ h = \rho e^{i\alpha}$ where the amplitude function $\rho$ is real, odd-Grassmann valued function of $u$ or $v$ whereas the phase function $\alpha$ is an angular function of $u$ or $v$.
 Plane waves correspond to the special choice $\alpha(u) = \kappa u$ or $\alpha(v) = \kappa v $ where
 $\kappa$ denotes the wave number.
Plane waves with constant amplitude give a subclass of solutions for which the neutrino field is non-zero but the metric functions remain blind
to the presence of a neutrino \cite{davis-ray2, hayward}..
Such configurations are called  the {\sl ghost neutrinos}. They are well-known in plane wave space-times \cite{collinson-morris,davis-ray}.
This naming was suggested in the first place because the neutrino source term on the right hand side of the Einstein equation that is identified as the symmetric energy-momentum
tensor of the neutrino vanishes.
 However, it was noted later that in the first order formalism where the  torsion contributes to the canonical energy-momentum, the ghost interpretation can no longer be maintained
  \cite{B14,B18-a,B15}.

\section{Concluding Remarks}

\noindent In conclusion, the study of neutrinos in general provide significant implications for both astrophysics and fundamental physics. By isolating the effects of gravity on neutrino trajectories and simplifying the equations describing neutrino interactions with matter, physicists can gain a deeper understanding of the physical processes that occur in astrophysical sources and test predictions of proposed extensions to the Standard Model. Furthermore, studying neutrinos in curved spacetimes may provide new avenues for detecting dark matter particles and shed light on some of the most pressing questions in physics today \cite{chianese-wu-king}. The potential for neutrinos to serve as probes of the universe's most energetic and exotic phenomena makes them a topic of continued interest and study in the scientific community.

Finally, the study of neutrinos in particular in plane symmetric spacetimes may provide new avenues for detecting dark matter particles. Dark matter, which is thought to make up the majority of the matter in the universe, has so far eluded detection through traditional observational methods. Dark matter interacts very weakly with light and other forms of matter, making its detection difficult. Observations of gravitational lensing and galactic rotation curves have provided indirect evidence for the existence of dark matter, but the nature of the dark matter particle remains unknown.
However , some theories propose that dark matter particles interact with neutrinos, which could lead to observable effects in the behavior of neutrinos. By studying neutrinos in plane symmetric spacetimes, physicists can better understand the possible interactions between dark matter and neutrinos, potentially providing new avenues for detecting dark matter particles.

\newpage

As a final remark we note that a considerable amount of the standard model's three-generation structure can be realized from the algebra of complex octonions \cite{furey,koivisto,todorov}.
This may be a promising  direction along which our study can be generalized\footnote{We thank our referee for pointing this out.}.

\section{Appendix}
We start with the identity
$
{}^*\bar{e} = \frac{1}{6} e^{*} \w e \w e^{*},
$
that implies
\ba
\nabla^*\bar{e}&=&\frac{i}{6}(\nabla e^* \w e \w e^* - e^* \w \nabla e \w e^* + e^* \w e \w \nabla e^*) \nonumber \\
&=&\frac{i}{6}(T^*\w e \w e^*- e^* \w T \w e^* +e^* \w e \w T^*)\nonumber \\
&=&\frac{i}{6}\mathcal{H}(2T^*\w e \w e^* - e^*\w T \w e^*). \nonumber
\ea
Then we use the variational field equations
$$
2T= 2 \mathcal{A}(h e^{*} \w e) - e \w \bar{h} \w e
$$
where $h = -\frac{1}{6} \xi \xi^{\dagger}$.  Then
\ba
\nabla {}^*\bar{e} = &=&\frac{i}{12}\mathcal{H}\Big( 2(h^*e \w e^* + e^*\w eh^* - e^*h\w e^*)\w e \w e^* \nonumber \\
&-& e^* \w (he^*\w e + e\w e^*h-eh^*\w e) \w e^* \Big)\nonumber \\
&=&\frac{i}{12}\mathcal{H}\Big(2h^*\w e \w e^*\w e \w e^*+3e^* \w eh^*\w e \w e^* - 4e^*\w he^* \w e \w e^*  \Big)\nonumber \\
&=&\frac{i}{12}\mathcal{H}\Big( 2(h^*e\w e^* + e^*\w eh^*)\w e \w e^* + e^* \w eh^* \w e \w e^* - 4e^*\w h e^*\w e \w e^* \Big). \nonumber
\ea
Now we use the identities
\ba
(h^*e \w e^*)^{\dagger}=(e\w e^*)^{\dagger}\bar{h}=-\bar{e}\w e^{\dagger}h^* = -e^* \w eh^*,
\ea
and
\ba
(e^* \w he^*)^{\dagger}=-(he^*)^{\dagger}\w \bar{e}=-\bar{e}h \w \bar{e}=-e^* \w he^*,
\ea
to simplify the above expression:
\ba
\nabla^*\bar{e}=\frac{i}{12}\mathcal{H}\Big( 4\mathcal{A}(h^*e\w e^*-e^*\w eh^*)\w e \w e^* +  e^* \w eh^* \w e \w e^*\Big).
\ea
The first term inside the big parantheses vanishes identically. The remaining term simplifies:
\ba
\nabla^*\bar{e}&=&\frac{i}{12}\mathcal{H}(e^* \w eh^* \w e \w e^*) \nonumber\\
&=&\frac{i}{12}(e^* \w eh^* \w e \w e^*)\nonumber \\
&=&-\frac{i}{72}(e^* \w e\xi)\w (\xi^{\dagger}e \w e^*) \nonumber \\
&=&\frac{i}{72}(e^* \w e\xi)\w (e^* \w e\xi)^{\dagger}.\ea
Now let us write $\xi= \alpha U^1+\beta U^2$ for some complex functions $\alpha$ and $\beta$.
After a long straightforward calculation one reaches, on-shell, the equality
\ba
\nabla^*\bar{e}=0.
\ea

\section{Acknowledgement}
One of us (T.D.) thanks the Turkish Academy of Sciences (TUBA) for partial support.
\section*{Data Availability Statement}
No new data were created or analysed in this study.


\bigskip

\begin{thebibliography}{99}
\bibitem{close} F.Close,{\bf Neutrino} (Oxford University Press,2010)
\bibitem{senjanovic} G.Senjanovic,{\sl Neutrino 2020:theory outlook},Int.J.Mod.Phys.{\bf A36}(2021)2130003
\bibitem{brill-wheeler} D.Brill and J.A.Wheeler,{\sl Interaction of neutrinos and gravitational fields}, Rev. Mod. Phys.{\bf 29} (1957) 465
\bibitem{wainwright} J.Wainwright,{\sl Geometric properties of neutrino fields in curved space-time}, J. Math. Phys. {\bf 12} (1971) 828
\bibitem{trim-wainwright} D.Trim and J.Wainwright,{\sl Combined neutrino-gravitational fields in general relativity}, J. Math. Phys.{\bf 12} (1971) 2494
\bibitem{griffiths1} J. B. Griffiths,{\sl Gravitational radiation and neutrinos}, Comm. Math. Phys. {\bf 28} (1972) 295
\bibitem{madore} J.Madore,{\sl On the neutrino in general relativity},Lett.Nuo.Cim.{\bf 5}(1972)48
\bibitem{taub} A.H.Taub, {\sl Empty space-times admitting a three parameter group of motions}, Ann.Math.{\bf 53}(1951)472
\bibitem{taub2} A.H.Taub, {\sl Isentropic hydrodynamics in plane symmetric space-times}.Phys.Rev.{\bf 103}(1956)454
\bibitem{taub3} A.H.Taub,{\sl Plane-symmetric similarity solutions for self-gravitating fluids} in {\bf General Relativity: Papers in Honour of J.J.Synge} Edited by L.O'Raifeartartaigh
(Oxford U.P.,1972) pp.133-150
\bibitem{carlson-safko} G.T.Carlson Jr.,J.L.Safko,{\sl An investigation of some of the kinematical aspects of plane symmetric space-times},J.Math.Phys.{\bf 19}(1978)1617
\bibitem{dereli-tucker} T. Dereli and R. W. Tucker,{\sl An intrinsic analysis of neutrino couplings to gravity}, J. Phys. {\bf A15} (1982) 1625
\bibitem{benn} I.M.Benn,{\sl  Complex quaternionic formulation of $SL(2,\mathbb{C})$ gauge theories of gravitation}.Unpublished Ph.D.thesis (Lancaster University,1981)
\bibitem{davis-ray2} T.M.Davis,J.R.Ray,{\sl Ghost neutrinos in plane-symmetric spacetimes},J.Math.Phys.{\bf 16}(1975)75
\bibitem{hayward} S.A.Hayward,{\sl Energy of gravitational radiation in plane-symmetric space-times},Phys.Rev.{\bf D78}(2008)044027
\bibitem{collinson-morris} C.D.Collinson,P.B.Morris,{\sl Spacetimes admitting neutrino fields  with zero energy and momentum},J.Phys.{\bf A6}(1973)915
\bibitem{davis-ray} T.M.Davis,J.R.Ray {\sl Ghost neutrinos in general relativity}, Phys.Rev.{\bf D9}(1974)334
\bibitem{B14} T.Dereli,R.W.Tucker {\sl Exact neutrino solutions in the presence of torsion},Phys.Lett.{\bf A 82},(1981)229
\bibitem{B18-a} J.B.Griffiths,{\sl Neutrino fields in Einstein-Cartan theory}, Gen.Rel.Grav.{\bf 13}(1981) 227
\bibitem{B15} A.Dimakis,F.M\"{u}ller-Hoissen,{\sl  Solutions of the Einstein-Cartan-Dirac equations with vanishing energy-momentum tensor},J.Math.Phys.{\bf 26}(1985)1040
\bibitem{chianese-wu-king} M.Chianese,B.Fu,S.F.King,{\sl Interplay between neutrino and gravity portals for FIMP dark matter}, JCAP {\bf 01}(2021)034
\bibitem{furey} C.Furey,{\sl Three generations,two unbroken gauge symmetries, and one eight-dimensional algebra},Phys.Lett.{\bf B785}(2018)84
\bibitem{koivisto} J.B.Jimenez,T.S.Koivisto,{\sl Listening to celestial algebras},Universe,{\bf 8}(2022)407
\bibitem{todorov} I.Todorov,{\sl Octonion internal space algebra for the standard model}, Universe,{\bf 9}(2023)222

\end{thebibliography}
\end{document}